\documentclass{ecai}
\usepackage{times}
\usepackage{graphicx}
\usepackage{latexsym}

\usepackage{amsmath}
\usepackage{booktabs}
\usepackage{amssymb}
\usepackage{marvosym}

\begin{document}

\title{ET-GAN: Cross-Language Emotion Transfer Based on Cycle-Consistent Generative Adversarial Networks}

\author{Xiaoqi Jia\textsuperscript{*} \and Jianwei Tai\textsuperscript{*} 
\institute{\textsuperscript{*} These authors contributed equally to this work.} 
\and Hang Zhou \and Yakai Li \and Weijuan Zhang \\ \and Haichao Du \and Qingjia Huang\textsuperscript{\Letter}
\institute{Key Laboratory of Network Assessment Technology, the Chinese Academy of Sciences, Beijing, China;
Beijing Key Laboratory of Network Security and Protection Technology, Beijing, China;
email: \{jiaxiaoqi,taijianwei,zhangweijuan,duhaichao,huangqingjia\}@iie.ac.cn. 
The Chinese University of Hong Kong, Hong Kong, China, email: zhouhang@link.cuhk.edu.hk.}
\institute{\textsuperscript{\Letter} Qingjia Huang is the corresponding author.}
}

\maketitle
\bibliographystyle{ecai}

\begin{abstract}
Despite the remarkable progress made in synthesizing emotional speech from text, it is still challenging to provide emotion information to existing speech segments. 
Previous methods mainly rely on parallel data, and few works have studied the generalization ability for one model to transfer emotion information across different languages.
To cope with such problems, we propose an emotion transfer system named ET-GAN, for learning language-independent emotion transfer from one emotion to another without parallel training samples. 
Based on cycle-consistent generative adversarial network, our method ensures the transfer of only emotion information across speeches with simple loss designs. 
Besides, we introduce an approach for migrating emotion information across different languages by using transfer learning.
The experiment results show that our method can efficiently generate high-quality emotional speech for any given emotion category, without aligned speech pairs. 
\end{abstract}

\section{Introduction}

Emotional speech plays a vital role in the field of human-computer interaction and robotics. As the naturalness of artificially synthesized speech is closer to that of real speech, the need to compensate for the deficiency of emotional expression in synthesized speech is becoming more obvious. Chiba \textit{et al.}~\cite{chiba2018an} point out that even in a non-task-oriented dialogue system, emotional speech can significantly improve user experience. Moreover, with the rapid development of brain-computer interface technology, synthesized emotional speech can help language-impaired people to express their emotions directly in the communication. These motivations have inspired researchers to synthesize emotional speech for given emotions in the last decade. Unfortunately, compared with other human emotion signals such as gestures, facial expressions, and postures, the emotion of speech is more complex and hard to regularize. To a certain extent, there is still a gap between synthesized emotional speech and real speech on authenticity and accuracy. 
In addition, most of the relevant research focuses on certain languages while ignoring the shared emotion representation among all languages. 
To cope with the problems mentioned above, a method for attaching emotion to cross-language speech is urgently needed.

The foundation for synthesizing more authentic emotional speech is to accurately extract the emotional features in phonetics. 
Traditionally, prosody features and spectrum features are the key components for analyzing emotional speech.
Based on speech spectrum and prosody modeling, desired emotions can be expressed in synthetic speech with the continuing research on accurate extraction of emotional features.
By changing the prosody, the emotion representation in speech can be changed obviously, and the adjustment of speech spectrum can affect the intensity of emotional representation. The methods based on Gaussian mixture model (GMM) \cite{kawanami2003gmm} and hidden Markov model (HMM) \cite{mustafa2013emotional} have been proposed to successfully synthesize emotional speech in experimental environments. Although the above works have achieved prospective results, due to the complexity of speech emotion, it is still difficult to achieve high-level control for emotional speech only by analyzing phonetics features.

Recently, tremendous progress has been made by applying deep learning to fit the mapping function for emotional information in high-dimensional space, which has a superior capability of regularizing the prosody of emotional speech than previous methods. 
In recent works \cite{an2017emotional,choi2019multi}, convolutional neural networks (CNNs) and recurrent neural networks (RNNs) have been proved to be effective for this task. Most methods are based on fine-tuning text-to-speech (TTS) systems or rely on a limited number of paired data.
However, both the TTS fine-tuning and the collection of paired data are cost-worthy. 
There is a lack of methods for independently transferring emotion information to speech without aligned data pairs.

In this paper, instead of learning to generate emotional speech in the TTS system, we focus on transferring specific emotions directly to existing speech without using paired data. To this end, we propose Emotion Transfer GAN (ET-GAN), a novel cross-language emotion transfer framework. 
We propose to base our network on cycle-consistent generative adversarial networks that can learn high-dimensional features between emotions. Non-parallel data can thus be effectively utilized. We then propose to balance various losses for identity preserving emotion transfer. Furthermore, to avoid gradient explosion/disappearance, we use the loss of the Earth-Mover (EM) distance \cite{arjovsky2017wasserstein} to replace the original loss of the model and add a gradient penalty \cite{gulrajani2017improved} term for stable convergence. 

It is noteworthy that deep learning models have potential cross-domain migration capabilities. Coutinho \textit{et al.}~\cite{coutinho2017shared} have demonstrated that shared acoustic models between speech and music have emotional commonality. Transfer learning provides a feasible way to alleviate the limitation of inadequate datasets and motivates the cross-language generalization of emotional speech synthesis model. Transfer learning technique is introduced in our work to improve the cross-language generalization of the model.

In the field of images, Fréchet Inception Distance (FID) \cite{lucic2018are} is widely used to evaluate the quality of generated images, while there are few similar metrics except the traditional ones with insufficient correlation with human perception on speech. Inspired by the latest research \cite{kilgour2018fr}, we first introduced Fréchet Audio Distance (FAD) with a high perception correlation to evaluate the quality of synthesized emotional speech. 
Experiments show that the proposed method can effectively synthesize high-quality emotional speech with excellent cross-language generalization performance. Through the above discussions, our contributions can be summarized as follows:

\begin{itemize}

\item We propose a novel parallel-data-free emotion transfer method named ET-GAN;

\item By introducing EM distance and gradient penalty, we provide a more stable training for the model without the issue of gradient explosion/disappearance;

\item We evaluate cross-language generalization of ET-GAN to simplify the migration of emotion transfer to any language by using transfer learning technique;

\item We adopt FAD as a speech evaluation metric to evaluate the performance of synthesized emotional speech with a high relevance to human perception.

\end{itemize}

\section{Related Works}
\noindent\textbf{Traditional Methods.}
In order to transform neutral speech into speech of different emotions, previous researchers focus on analyzing the prosody and spectrum features of emotional speech. Kawanami \textit{et al.}~\cite{kawanami2003gmm} proposed GMM-based emotional speech synthesis method to synthesize speech emotion under specific emotion categories. Tao \textit{et al.}~\cite{tao2006prosody} proposed a pitch-targeting model for describing the F0 contour of Mandarin, which evaluates the expressiveness of synthesized speech emotion by the deviation of perceptual expressiveness (DPE). Unlike specific emotion categories, the pitch-targeting model synthesizes emotional speech based on three intensity levels. Wen \textit{et al.}~\cite{wen2011prosody} designed a tone nucleus model to represent and transform F0 contours for synthesizing emotional speech at a comparable level to that of professional speakers by just using a small amount of data. As an improvement of GMM-based emotional speech synthesizer, \cite{aihara2012gmm} obtained more expressive emotional speech by combining spectrum and prosody features. \cite{mustafa2013emotional} proposed a Malay emotional speech synthesis system based on HMM, which includes deterministic annealing expectation maximization algorithm and automatic segmentation seed model. To improve the accuracy of acoustic feature estimation and rules for modification, \cite{xue2018voice} used fuzzy inference system (FIS) to correlate acoustic features, emotion dimensions (valence and activation) and semantic primitives and used Fjuisaki model and target prediction model to parameterize prosody-related features. Nevertheless, the regularization of emotional representation in speech is still a serious challenge.

\noindent{\textbf{TTS-based Methods.}} Many research works \cite{potard2016cross,kalita2017emotional} also discussed how to fine-tune the existing Text-to-Speech System (TTS) by appropriately allocating prosodic parameters for increasing emotional expression in the output speech. Emotional speech synthesis method proposed by \cite{lee2017emotional} and \cite{skerry-ryan2018towards} has achieved impressive success through attaching extension modules for Tacotron which is an advanced TTS system. 

\noindent{\textbf{Deep Learning-based Methods.}} 
In recent years, emotional speech synthesis based on deep learning has become an emergent area of interest. LSTM-based emotional statistical parametric speech synthesis method is proposed by \cite{an2017emotional}, and \cite{li2018emphasis} further tried to synthesize questionable and exclamatory speech rather than specific emotional categories with a real time synthesis system. 

With the rapid development of deep learning, convolutional neural networks been proven effective in the field of audio processing. For example, CNNs are used for speaker recognition and verification~\cite{nagrani2017voxceleb, chung2018voxceleb2}, extracting information for video generation~\cite{chung2017you,zhou2019talking} and audio generation~\cite{marafioti2019adversarial}. Particularly, voice conversion~\cite{kaneko2018cyclegan,kameoka2018stargan,kaneko2019cyclegan} is more related to emotion transfer. 

Choi \textit{et al.}~\cite{choi2019multi} combined speaker information and emotional expression based on CNN by encoding speakers and emotions. However, the methods based on deep learning rely on a large number of labeled and paired data, which is difficult to obtain. In addition, few studies have focused on the model's cross-language generalization performance, although it is an obvious problem to be solved for the rapid development of software services.

\begin{figure*}
  \centering
  \includegraphics[width=0.9\textwidth]{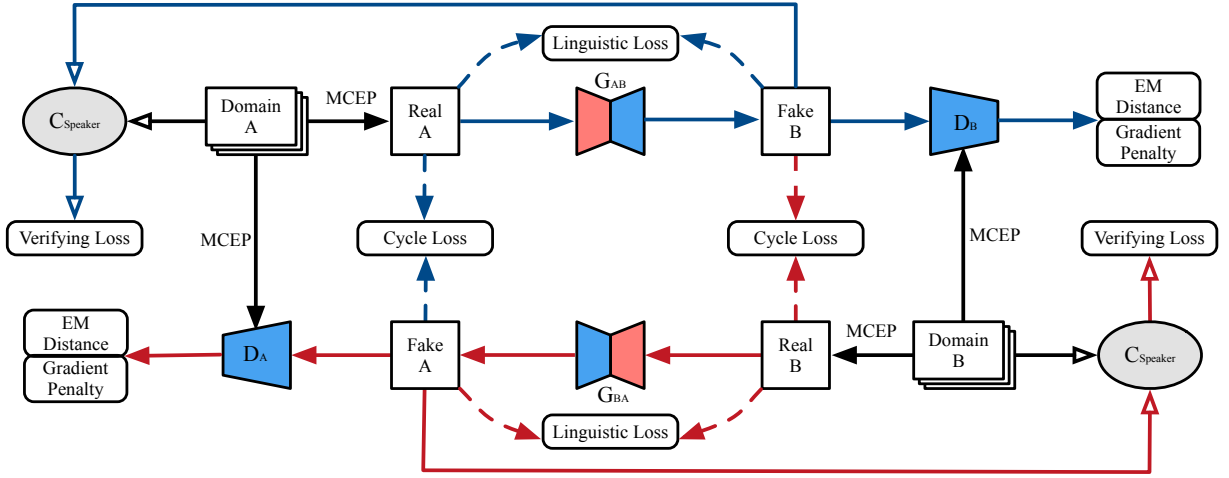} 
  \caption{ET-GAN Architecture. Real audio clips Real A and Real B are extracted from different emotion domains A and B respectively. The cycle loss represents the whole cycle-consistency adversarial training. Linguistic loss, speaker verifying loss and gradient penalty are added symmetrically to this pipeline.} %
  \label{ET-GAN} %
\end{figure*}

\section{Our Method}
\label{model}

\subsection{CycleGAN}

We start by briefly reviewing the concept and formulation of CycleGAN \cite{zhu2017unpaired}. CycleGAN learns a mapping function between two domains $X$ and $Y$ without any set of aligned image pairs. CycleGAN includes two mappings $G : X \rightarrow Y$ and $G : Y \rightarrow X$, which are learned by using two different losses, namely the adversarial loss \cite{goodfellow2014generative} and the cycle-consistency loss with two different discriminators. An adversarial loss is used to measure how distinguishable converted data $G_{X\rightarrow Y}\left (x \right )$ are from target data $y$, where $x\in X$ and $y\in Y$. Therefore, the closer the distributions of converted data $P_{G_{X\rightarrow Y}}\left (x \right )$ and target data $P_{data}\left (y \right )$ become, the smaller the adversarial loss will be. This adversarial loss function is written as 

\[
\begin{aligned}
L_{adv}\left ( G_{X\rightarrow Y},D_{Y},X,Y \right ) = \mathbb{E}_{y\sim P_{data}\left (y \right )}\left [ log D_{Y} \left ( y \right )\right ] \\ 
+ \mathbb{E}_{x \sim P_{data}\left (x \right )}\left [ \log\left ( 1 - D_{Y} \left ( G_{X\rightarrow Y}\left (x \right ) \right )\right )\right ] 
\end{aligned} 
\eqno{(1)}
\]

where generator $G_{X\rightarrow Y}$ attempts to minimize this loss to generate speech samples indistinguishable from target data $y$ by the discriminator $D_{Y}$, while $D_{Y}$ aims to maximize this loss to distinguish between translated speech samples $G_{X\rightarrow Y}$ and real samples $y\in Y$. As described above, the formula can be expressed as $\min_{G_{X\rightarrow Y}}\max_{D_{Y}}L_{GAN}\left ( G_{X\rightarrow Y},D_{Y},X,Y \right ) $. In CycleGAN, there is a similar adversarial loss for the mapping function $G : Y \rightarrow X$ with its discriminator $D_{X}$, and the objective is shown as $\min_{G_{Y\rightarrow X}}\max_{D_{X}}L_{GAN}\left ( G_{Y\rightarrow X},D_{X},Y,X \right ) $ \cite{zhu2017unpaired}.

Theoretically, adversarial training has the ability to learn stochastic functions $G$ that produce outputs identically distributed as target domain $Y$ and source domain $X$ respectively. Unfortunately, when the capacity is large enough, the network can map the same set of samples in the source domain to randomly arranged samples in the target domain, where any known mapping can induce an output distribution that matches the target distribution \cite{zhu2017unpaired}. The context information of $X$ and $G_{X\rightarrow Y}$ are not necessarily consistent only by optimizing the adversarial loss. Because the adversarial loss only confirms that $G_{X\rightarrow Y}\left (x \right )$ follows the target data distribution, and do not help save the context information of $x$. In order to further reduce the possible mapping function space, the mapping function learned by the model should be circularly consistent, in detail, $G_{Y\rightarrow X}\left (G_{X\rightarrow Y}\left (x \right ) \right )\approx x$ and $G_{X\rightarrow Y}\left (G_{Y\rightarrow X}\left (y \right ) \right )\approx y$. In CycleGAN, two additional items are introduced to address this problem. One is an adversarial loss $L_{GAN}\left ( G_{Y\rightarrow X},D_{X},Y,X \right )$ for an inverse mapping $G_{Y\rightarrow X}$ and the other is a cycle-consistency loss, written as

\[
\begin{aligned}
& L_{cyc}\left ( G_{X\rightarrow Y},G_{Y\rightarrow X} \right) \\
&\quad \quad = \mathbb{E}_{x\sim P_{data}\left (x \right )}\left [\left \| G_{Y\rightarrow X}\left (G_{X\rightarrow Y}\left (x \right ) \right ) - x \right \|_{1} \right ] \\
&\quad \quad + \mathbb{E}_{y \sim P_{data}\left (y \right )}\left [ \left \| G_{X\rightarrow Y}\left (G_{Y\rightarrow X}\left (y \right ) \right ) - y \right \|_{1} \right ].
\end{aligned}
\eqno{(2)}
\]

The cycle-consistency loss makes $G_{X\rightarrow Y}$ and $G_{Y\rightarrow X}$ to seek out $\left (x, y\right )$ pairs which have the same context information as far as possible. With weighted parameter $\lambda _{cyc}$, the full objective of CycleGAN \cite{zhu2017unpaired} can be summarized as

\[
\begin{aligned}
L\left ( G_{X\rightarrow Y},G_{Y\rightarrow X} \right) 
&=L_{adv}\left ( G_{X\rightarrow Y},D_{Y},X,Y \right ) \\
&+L_{adv}\left ( G_{Y\rightarrow X},D_{X},Y,X \right ) \\
&+L_{cyc}\left ( G_{X\rightarrow Y},G_{Y\rightarrow X} \right ).
\end{aligned}
\eqno{(3)}
\]

\subsection{ET-GAN}


Our goal is to learn a language-independent mapping between different emotion domains without aligned speech pairs. To achieve this goal, we implement ET-GAN base on CycleGAN \cite{zhu2017unpaired}. However, directly applying CycleGAN can lead to the consequence of achieving voice conversion. So in the ET-GAN architecture, we make three modifications to the CycleGAN architecture to adopt CycleGAN to our task: linguistic-information loss, speaker-verifying loss and gradient penalty. Figure \ref{ET-GAN} illustrates the architecture of our proposed model. 

\paragraph{Linguistic-information loss}
Although the loss of circular consistency provides constraints on the structure of learned mappings, such constraints are insufficient to ensure that learned mappings always retain linguistic information. Inspired by the loss of identity mapping \cite{taigman2017unsupervised}, we introduce the linguistic-information loss to retain linguistic information of learned mappings. Linguistic-information loss can encourage generators to learn mappings that preserve combinations between inputs and outputs. According to the definition of identity mapping, linguistic-information loss can be written as

\[
\begin{aligned}
L_{li}\left ( G_{X\rightarrow Y},G_{Y\rightarrow X} \right) & = \mathbb{E}_{x\sim P_{data}\left (x \right )}\left [\left \| G_{X\rightarrow Y}\left (x \right )  - x \right \|_{1} \right ] \\
&+ \mathbb{E}_{y \sim P_{data}\left (y \right )}\left [ \left \| G_{Y\rightarrow X}\left (y \right ) - y \right \|_{1} \right ].
\end{aligned}
\eqno{(4)}
\]

\paragraph{Speaker-verifying loss}
Recognition of a person by voice is an important human feature, and recognizing the speaker's identity is an important prerequisite for continuing communication. Therefore, a familiar speaker's voice will make people more accustomed to listening to the speech. We aim to generate pure emotion information and to change the speaker's phonological features as little as possible. In other words, transferring emotion to speech through our model will not change the speaker verification features. In detail, we introduce a speaker feature extraction module with a discriminator to encode and verify the speaker. The speaker-verifying loss constrains the range of speaker features to ensure that the generated target speech has the same speaker attributes as the original speech. Inspired by VGGVox \cite{nagrani2017voxceleb}, we implement feature extraction and verification of speakers based on this model. Since speaker identification is considered as normal classification task \cite{nagrani2017voxceleb}, the output of the feature extraction module is fed into a softmax to produce a distribution over different speakers. The prediction of speaker identification is represented as $p_{i}$ and the value of real sample is $t_{i}$. In our method, the speaker-verifying loss can be simply written as

\[
\begin{aligned}
L_{sv}\left ( G_{X\rightarrow Y},G_{Y\rightarrow X} \right) 
=-\sum t_{i} \log p_{i}.
\end{aligned}
\eqno{(5)}
\]

\paragraph{Gradient penalty}
During GAN training process, the distribution of generated data is encouraged to approximate the distribution of real data. The objective function of GAN consists of all F divergences and exotic combinations. The traditional GAN always faces the problem of training instability, so we need to pay attention to the structure of the model. Besides the training of generator and discriminator must be carefully coordinated. To address this problem, the Earth-Mover (EM) distance is introduced to replace the original loss measurement method while the weight clipping method is proposed to implement Lipschitz constraint indirectly \cite{arjovsky2017wasserstein}. At the same time, Martin \textit{et al.}'s subsequent work \cite{gulrajani2017improved} theoretically explained the EM distance and compared it to other commonly used distance and divergence formulas (such as JS divergence and KL divergence), and the results prove the EM distance is still continuous and can be derived in low dimensional space. The EM distance is continuous and the gradient is decreasing while the JS is discontinuous. Since the state distribution is continuous, discrete evaluation cannot be used. In our training process, the EM distance is used to replace the traditional loss method to solve the problem of gradient explosion/disappearance in the original GAN, which simplifies the adjustment of the generator and discriminator during the adversarial training process.
However, weight clipping limits the range of values of each network parameter independently, so the optimal strategy will make all parameters go to extremes as far as possible. Therefore gradient penalty \cite{gulrajani2017improved} is proposed to implement Lipschitz constraint as an optimization scheme for weight clipping to avoid undesired behavior. Due to the fact that Lipschitz constraint requires the gradient of discriminator not to exceed constant K, we introduce gradient penalty into ET-GAN to implement Lipschitz constraint with an additional loss term which establishes the relationship between the gradient and K. Gradient penalty will encourage more stable convergence of ET-GAN and circumvent gradient explosion/disappearance issues. Assuming that the random sample is represented as $P_{data}\left (\widehat{x} \right )$. ET-GAN is based on cycle-consistent adversarial networks with the gradient penalty. The adversarial loss function of ET-GAN with the weight parameter $\lambda _{gradient}$ is written as

\[
\begin{aligned}
&L_{adv}\left ( G_{X\rightarrow Y},D_{Y},X,Y \right) \\
&=\mathbb{E}_{x\sim P_{data}\left (x \right )}\left [ D_{Y} \left (  G_{X\rightarrow Y}\left (x \right )  \right )  \right ]
- \mathbb{E}_{y\sim P_{data}\left (y \right )}\left [ D_{Y}\left (y  \right )  \right ] \\
&+ \lambda _{gradient} \mathbb{E}_{\widehat{x}\sim P_{data}\left (\widehat{x} \right )}\left [\ \left (   \left \| \bigtriangledown _{\widehat{x}}D_{Y}\left ( \widehat{x} \right ) \right \|_{2} - 1 \right )^{2} \right ].
\end{aligned}
\eqno{(6)}
\]

ET-GAN is constrained by three additional loss functions: linguistic-information loss, speaker-verifying loss and gradient penalty loss. Hence, our full objective is

\[
\begin{aligned}
&L_{full}\left ( G_{X\rightarrow Y},G_{Y\rightarrow X} \right)
=L_{adv}\left ( G_{X\rightarrow Y},D_{Y},X,Y \right)\\
&+L_{adv}\left ( G_{Y\rightarrow X},D_{X},Y,X \right) 
+\lambda _{cyc} L_{cyc}\left ( G_{X\rightarrow Y},G_{Y\rightarrow X} \right)\\
&+\lambda _{li} L_{li}\left ( G_{X\rightarrow Y},G_{Y\rightarrow X} \right)
+\lambda _{sv} L_{sv}\left ( G_{X\rightarrow Y},G_{Y\rightarrow X} \right).
\end{aligned}
\eqno{(7)}
\]

\subsection{Transfer Learning}

In our proposed scenario, it is assumed that each domain $D$ consists of feature space $Z$ and edge probability distribution $P\left ( S \right )$, where the sample $S=s_{1},s_{2},...,s_{n} \in Z$. Given a domain $D= \left \{ Z, P\left ( X \right ) \right \}$, a task $T$ is composed of the emotion label space $E$ and the mapping function $f$. Because emotions are common in different languages, the emotion information extracted from one language can be transferred to another language only through a small amount of training and data theoretically. In the task of generating emotion, the task domain $T_{a}$, $T_{b}$ based on different languages $A$ and $B$ can be expressed as $D_{a}$ and $D_{b}$. $P_{a}\left ( E | S =s \right ) = P_{b}\left ( E | S =s \right )$ holds for $\forall s \in Z$, and $P_{a}\left ( S \right )\neq P_{b}\left ( S \right )$. The purpose of transfer learning is to map data from different domains (e.g. $D_{a}$, $D_{b}$) into an emotion space $E$, and make it as close as possible in the space. Hence, the mapping function trained by the source domain $D_{a}$ in emotion space $E$ can then be migrated to the target domain $D_{b}$ to improve the accuracy of the mapping function in the target domain $D_{b}$. By introducing transfer learning, ET-GAN can share the knowledge of emotion information in different languages by sharing a part of model parameters, thus implement an excellent cross-domain generalization performance with feature representation transfer. Furthermore, transfer learning can alleviate the lack of labeled emotional speech samples.

\section{Experiments}

\subsection{Datasets}

We conducted our experiments to evaluate ET-GAN on emotion transfer tasks without aligned speech pairs. Four different datasets are used in our experiments. The Interactive Emotional Dyadic Motion Capture Database (IEMOCAP) \cite{busso2008iemocap}, Berlin Database of Emotional Speech (Emo-DB) \cite{burkhardt2005a}, Canadian French Emotional Speech Dataset (CaFE) \cite{gournay2018a} and Mandarin Emotional Speech Dataset (MEmoSD) provided by our research team. MEmoSD is recorded in a laboratory environment by four professional Mandarin speakers (2 males and 2 females) according to four emotion categories ($Angry$, $Happy$, $Neutral$, $Sad$). To maintain uniformity in emotion categories, a subset is selected from IEMOCAP, Emo-DB and CaFE in our experiments: $Angry$, $Happy$, $Neutral$ and $Sad$. We utilized MEmoSD and IEMOCAP to validate the performance of ET-GAN, while we use Emo-DB and CaFE to further evaluate the cross-language generalization performance of ET-GAN with transfer learning. In the preprocessing, all speech inputs are sampled at 16kHz or 48kHz, and 24-dimension Mel-cepstral coefficients (MCEPs) are extracted by using the WORLD analysis system \cite{morise2016world}.

\subsection{Network Architectures}

ET-GAN adopts the network architecture from CycleGAN-VC \cite{kaneko2018cyclegan,kaneko2019cyclegan}. To capture the wide-range temporal structure while preserving the information of sample structure, the generator networks of ET-GAN are designed as a 1-dimension(1D) CNN with several residual blocks, including 2 downsample blocks, 6 residual blocks and 2 upsample blocks. A gated CNN \cite{dauphin2016language} is used in the generator to extract the acoustic features from time sequences and hierarchical structures. Due to the outstanding results of high-resolution image generation \cite{shi2016real}, we use pixel shuffler in upsampling blocks. For the discriminator networks, we use a 2-dimension (2D) CNN to handle a 2D spectral texture. To make a more stable convergence of the generator networks, the Earth-Mover (EM) distance is used in the discriminator which can distinguish real samples from fake samples to measure the discriminator loss. Meanwhile, we add a gradient penalty loss into the discriminator to prevent gradient explosion/disappearance issues by implementing the Lipschitz constraint. It's worth mentioning that instance normalization is used in both the generator and the discriminator to normalize a single sample in a batch. An auxiliary classifier is designed based on VGGnet \cite{nagrani2017voxceleb} to extract, encode and classify speaker features, which retains unchanged speaker features during the process of emotion domain transfer.

\subsection{Training}

In our experiments, we use three Nvidia RTX2070 graphics cards. MCEPs of all speech samples are extracted and normalized to unit variance and zero mean during preprocessing. We randomly cut 128-frame segment from speech samples with different lengths to ensure the randomness of the training process. The network is trained through Adam optimizer with $\beta _{1} = 0.5$ and $\beta _{2} = 0.9$. We set the generator's learning rate to 0.0002, the discriminator's learning rate to 0.0001, and the auxiliary speaker classifier's learning rate to 0.0001. The weight parameters are set as $\lambda _{cyc} = 10$, $\lambda _{li} = 5$, $\lambda _{sv} = 1$ and $\lambda _{gradient} = 5$. For migration learning on datasets of different languages, except the classification layer, all layers of the model are migrated to maximize the shared features of emotion information by using transfer learning technique. TTS samples provided by the Amazon Polly \cite{aws} and the Iflytek Open Platform \cite{xunfei} are used as test sets in all experiments and all TTS sample sets contain more than 3 speakers with different genders. Note that no additional process is used to align speech samples, which ensures that the experiments are not based on a training set with aligned speech pairs.

\begin{table*}
\caption{FAD and naturalness MOS with 95$\%$ confidence intervals}
\label{FAD_MOS}
\centering
\vspace{-5pt}
\begin{tabular}{@{}llcccclcccc@{}}
\toprule
        &     & \multicolumn{4}{c}{IEMOCAP}                             &           & \multicolumn{4}{c}{MEmoSD}                              \\ \cmidrule(lr){3-6} \cmidrule(l){8-11} 
        Metric    & Method                   & Angry        & Happy        & Sad          & Neutral    &           & Angry        & Happy        & Sad          & Neutral    \\ \midrule
\multicolumn{1}{l|}{FAD}      & StarGAN         & 5.86          & 5.49          & 5.35          & 4.55        &           & 1.88          & 3.11          & 2.53          &          2.25        \\
\multicolumn{1}{l|}{}           & CycleGAN        & 4.40          & 4.56          & 4.82          & 3.21        &           & 1.05          & 2.45          & 1.27          & 1.65        \\
\multicolumn{1}{l|}{}           & CycleGAN-improved        & 4.29          & 4.13          & 4.29          & 3.09        &           & 1.17          & 1.47          & 1.02          & 0.83        \\
\multicolumn{1}{l|}{}           & \textbf{ET-GAN}             & 4.25          & 4.14          & 4.34          & 3.05        &           & \textbf{0.58}          & {0.77}   & {0.99}          & {0.82}        \\
\multicolumn{1}{l|}{} & \textbf{ET-GAN-TL} & \textbf{3.88} & \textbf{3.21} & \textbf{3.78} & \textbf{2.77} &           & {0.64} & \textbf{0.70} & \textbf{0.91} & \textbf{0.72} \\ \midrule
\midrule
\multicolumn{1}{l|}{MOS}        & StarGAN         & 0.58          & 0.75          & 0.62          & 0.95        &           & 1.85          & 1.75          & 1.92          & 2.10        \\
\multicolumn{1}{l|}{}          & CycleGAN        & 2.12          & 1.20          & 1.72          & 1.89        &           & 2.88          & 2.30          & 2.78          & 2.84        \\
\multicolumn{1}{l|}{}          & CycleGAN-improved        & 2.69          & 2.77          & 2.65          & 2.59        &           & 3.85          & 3.38          & 3.65          & 3.52        \\
\multicolumn{1}{l|}{}          & \textbf{ET-GAN}             & 2.75          & 2.72          & 2.62          & 2.88        &           & \textbf{4.12}          & {3.62}   & {3.58}          & {3.75}        \\
\multicolumn{1}{l|}{}     & \textbf{ET-GAN-TL} & \textbf{2.81} & \textbf{2.92} & \textbf{2.84} & \textbf{2.95} &           & {4.05} & \textbf{3.80} & \textbf{3.64} & \textbf{3.92} \\ \bottomrule
\end{tabular}
\vspace{20pt}
\end{table*}

\begin{figure*}
  \centering
  \includegraphics[width=0.89\textwidth]{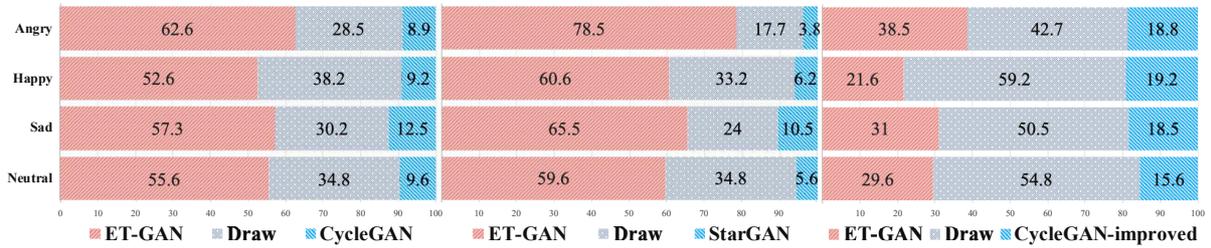} 
  \caption{Average preference proportion ($\%$) on emotion correctness} 
  \label{emotion_mos} 
\end{figure*}

\begin{figure*}
  \centering
  \includegraphics[width=0.89\textwidth]{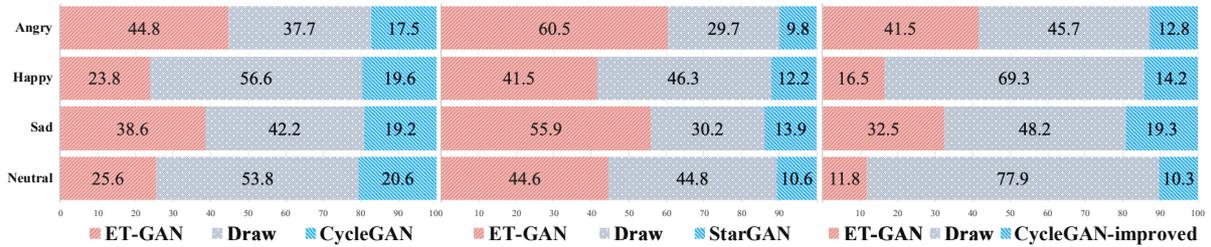} 
  \caption{Average preference proportion ($\%$) on speaker similarity} 
  \label{speaker_mos}
\end{figure*}

\begin{figure*}
  \centering
  \includegraphics[width=0.89\textwidth]{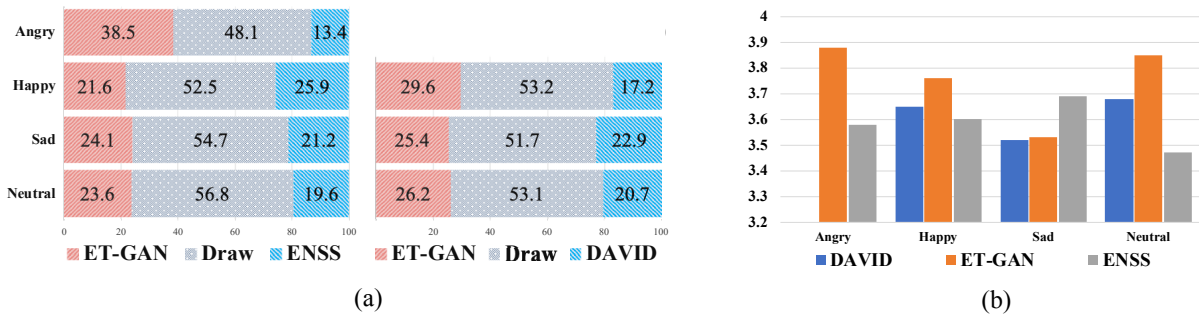} 
  \caption{(a) Average preference proportion ($\%$) on emotion correctness, (b) MOS with 95$\%$ confidence intervals.} 
  \label{new_mos}
\end{figure*}

\subsection{Objective Evaluation}

It is fairly intricate to use a single metric to comprehensively evaluate the quality of MCEPS after transfer. Therefore, the Mel-cepstral distortion (MCD) is used to evaluate the difference in the global structure. Also, the modulation spectra distance (MSD) is used to evaluate the difference of local structure in the follow-up work of \cite{kaneko2019cyclegan}. Although using two metrics to measure the quality of MCEPS is a feasible solution, in speech generation tasks, we want to directly evaluate the quality of synthesized speech rather than use MCEPS as an intermediate result. On the other hand, the objective metric is expected to have a higher correlation with human perception, in order to make the metric more instructive. 

Thanks to FAD \cite{kilgour2018fr} proposed by Kilgour \textit{et al.}, it is possible to solve the above problems. More specifically, unlike existing audio evaluation metrics, FAD evaluates the distribution between the generated samples and the real samples from the features extracted by a VGGish \cite{hershey2017cnn} model which is trained on a large clean set of audio. Compared with traditional evaluation metrics (distortion ratio, cosine distance, and magnitude L2 distance), FAD can accurately evaluate the effect of synthesized speech and has a higher correlation with human perception as a reference-free evaluation metric \cite{kilgour2018fr}. FAD has been innovatively introduced into our experiments for the first time to evaluate the quality of synthesized emotional speech. For FAD metrics, smaller values represent that the synthesized speech's emotion is more similar to the real target speech's emotion. We choose three advanced domain transfer models as the baseline methods in our experiments: StarGAN \cite{kameoka2018stargan}, CycleGAN \cite{kaneko2018cyclegan}, CycleGAN-improved \cite{kaneko2019cyclegan}. During the experiment, it is noteworthy that we have not used any pre-training weights and parameters on baselines, and all baselines were retrained in the emotion transfer paradigm. At the same time, the datasets, data preprocessing and training iterations are consistent with our proposed method, so as to reduce the interference of unrelated factors on experimental results.

Table \ref{FAD_MOS} shows the FAD of synthesized emotional speech in four targeted emotion domains over dataset IEMOCAP and MEmoSD. Compared with the baseline methods, Table \ref{FAD_MOS} proves that ET-GAN can synthesize higher quality emotional speech under limited iterations of the generator. It can be noted that the FADs on IEMOCAP is higher than that of MEmoSD. The reason is that the IEMOCAP dataset contains a lot of noise while MEmoSD is recorded in the lab environments. Thus the performance is better with cleaner data.

\subsection{Subjective Evaluation}

We take a mean opinion score (MOS) as a subjective evaluation metric. A naturalness MOS test (terrible: 0 to outstanding: 5) is used to measure the naturalness of synthesized emotional speech. To verify the emotion correctness of synthesized emotional speech, we use an XYR MOS test, where ``X'' and ``Y'' are the emotional speech synthesized from the baseline methods and the proposed method respectively, while ``R'' is the real emotional speech in the target domain. Synthesized emotional speech is randomly selected and played to the listener in random order to avoid the selection from sequential inertia. The listeners are asked to choose one (``X'' or ``Y'') that is more similar to the emotion of the real speech ``R'' after playing ``X'' and ``Y''. ``Draw'' can be chosen when the listener is too tough to make a clear choice. Similar to this approach, we use XYR MOS test to measure the speaker's similarity between synthetic emotional speech and original speech, where ``R'' represents the original speech. In the naturalness MOS test and XYR MOS test, 20 sentences for each emotion category are randomly selected from synthesized emotional speech. 30 listeners (15 males and 15 females) with good language proficiency are invited to participate in our experiment. For all MOS metrics, higher MOS values represent better performance of synthesized emotional speech in corresponding aspects. We ensure that the test sets for each emotion category are the same in the textual content of speech samples. In subjective experiments, speech samples that we play to the listener for each round of testing are generated by the same input speech, so the text content of them is the same. We minimize the misleading by other unrelated speech attributes (e.g. text, background sounds) to ensure the rigor and reliability of the experimental results. In the subjective evaluation, we supplemented the comparative experiment between ET-GAN and existing emotional speech synthesis methods. We chose two state-of-the-art emotional speech synthesis methods to perform a subjective test (MOS and emotion correctness preference): Emotional End-to-End Neural Speech synthesizer (ENSS) \cite{lee2017emotional} and DAVID \cite{rachman2018david}. Since DAVID does not include the emotional category of $Angry$, we only evaluate the speech samples of three emotions: $Happy$, $Neutral$, $Sad$.

Table \ref{FAD_MOS} shows the naturalness MOS test results of emotional speech synthesized by different methods in the 200th generator iteration. It is obvious that the emotional speech synthesized by ET-GAN is better than other baseline methods in naturalness. Emotional speech synthesized are evaluated by XYR MOS test for emotion correctness and speaker similarity, which are shown in Figure \ref{emotion_mos} and Figure \ref{speaker_mos}. Compared with other baseline methods, Figure \ref{emotion_mos} illustrates the subjective preference of listeners for the emotional speech synthesized by ET-GAN. And Figure \ref{speaker_mos} proves that ET-GAN has a slightly superior performance over the task of maintaining the speaker representation. The experimental results demonstrate that ET-GAN has better performance in emotion correctness and speaker similarity with a high level of control of emotion information. Results of the comparative experiment between ET-GAN, ENSS and DAVID are illustrated in Figure \ref{new_mos}, and prove that ET-GAN has higher MOS scores than other state-of-the-art emotional speech synthesis methods.

\begin{table*}
\caption{Transfer Learning}
\label{adaptation}
\centering{}
\vspace{-5pt}
\begin{tabular}{@{}lccccclcccc@{}}
\toprule
      &    & \multicolumn{4}{c}{FAD}                             &           & \multicolumn{4}{c}{MOS} \\                              \cmidrule(lr){3-6} \cmidrule(l){8-11} 
Method     &Language          & Angry        & Happy        & Sad          & Neutral    &           & Angry        & Happy        & Sad          & Neutral    \\ \midrule
ET-GAN     &German         & 2.73          & 2.85          & 2.96          & 2.64        &           & 3.70          & 3.36          & 3.15          & 3.45        \\
\textbf{ET-GAN-TL}  &German        & \textbf{2.28}          & \textbf{2.45}          & \textbf{2.37}          & \textbf{2.19}        &           & \textbf{3.85}          & \textbf{3.55}          & \textbf{3.24}          & \textbf{3.66}        \\
\midrule
ET-GAN     &French         & 3.24          & 3.09          & 3.64          & 3.71        &           & 3.15          & 3.20          & 2.85          & 2.98        \\
\textbf{ET-GAN-TL}  &French         & \textbf{3.02}          & \textbf{2.77}          & \textbf{3.34}          & \textbf{3.20}        &           & \textbf{3.40}          & \textbf{3.32}          & \textbf{2.90}          & \textbf{3.12}        \\
\bottomrule 
\end{tabular}
\vspace{+10pt}
\end{table*}

\begin{table*}
\caption{Ablation Study}
\label{ablation}
\centering
\vspace{-5pt}
\begin{tabular}{@{}lcccclcccc@{}}
\toprule
         & \multicolumn{4}{c}{FAD}                             &           & \multicolumn{4}{c}{MOS} \\                              \cmidrule(lr){2-5} \cmidrule(l){7-10} 
Method          & Angry        & Happy        & Sad          & Neutral    &           & Angry        & Happy        & Sad          & Neutral    \\ \midrule
 Cycle         & 3.98          & 3.19          & 3.68          & 3.75        &           & 0.78          & 0.80          & 0.60          &          0.75        \\
 Cycle+Linguistics        & 1.09          & 1.52          & 1.13          & 2.15        &           & 2.80          & 2.24          & 2.95          & 2.05        \\
 Cycle+Speaker         & 4.02          & 3.11          & 3.89          & 2.99        &           & 0.75          & 0.85          & 0.54          & 0.90        \\
 Cycle+Linguistics+Speaker         & 0.76          & 1.44          & 1.05          & 1.87        &           & 3.30          & 2.95          & 3.05          & 3.10        \\
 Cycle+Linguistics+Speaker+Gradient Clip       & 0.72          & 0.79          & 1.02          & 1.79        &           & 3.90          & 3.55          & 3.50          & 3.46        \\
\textbf{ET-GAN(ours)} & \textbf{0.58} & \textbf{0.77} & \textbf{0.99} & \textbf{0.82} &           & \textbf{4.12} & \textbf{3.62} & \textbf{3.58} & \textbf{3.75} \\  \bottomrule 
\end{tabular}
\vspace{15pt}
\end{table*}

\subsection{Transfer Learning}

In order to measure the effect of emotion transfer in different languages intuitively, we compared the effects of direct training model and model trained from pre-training model of Mandarin (MEmoSD) transfer by using transfer learning techniques on German (Emo-DB) and French (CaFE). According to objective evaluation results, under limited iterations, using domain adaptive technology can significantly improve the quality of synthesized emotional speech. This advantage proves that the emotion learned by ET-GAN is common and transferable between different languages. In view of reasonable inference, the proposed method can be expected to be able to extend to any other languages rapidly and efficiently. 

As for subjective evaluation, in the 200th generator iterations, Table \ref{adaptation} shows the naturalness MOS of the synthesized emotional speech, which proves that the performance of the model using transfer learning technique is better than that of the model not used.

\section{Ablation Study}

To further evaluate the effectiveness of each module we proposed, a comprehensive ablation study has been conducted as illustrated in Table \ref{ablation}. The ablation study is carried out on MEmoSD dataset for all the emotions. Both FAD and MOS are evaluated.

ET-GAN consists of four modules: the cycleGAN backbone, the linguistic-information loss $L_{li}$, the speaker-verifying loss $L_{sv}$, and gradient penalty. The cycleGAN backbone cannot be removed, thus serves as a baseline. Partial variants of adding each component to our baseline are trained. Besides, to validate the priority of our gradient penalty, an alternative version of gradient clipping is also evaluated. The training process and all hyper-parameters are kept the same as our proposed one.

A conclusion can be drawn according to Table \ref{ablation}. Linguistic loss which restricts the distances between the spectrum is crucial for preserving the content of speech, thus adding it improves the baseline by a large margin. However, only by adding the speaker verification loss, the model tends to be difficult to converge without controlling linguistic information. The loss of semantic information leads to bad performance. However, by combining both the linguistic and the speaker loss, the network can leverage the advantages of each of them and lead to superior results. Further results show that GAN is able to perform with gradient clipping in a steadier way. But our proposed gradient penalty outperforms that of gradient clipping. Thus verified the effectiveness of gradient penalty. 
Overall, our final model performs the best by means of both objective and subjective. And each component of our proposed model can be coordinated to work together towards a stable and effective emotion transfer framework. Each part of them contributes to our final performance.

\section{Discussion and Conclusion}

We propose a novel parallel-data-free cross-language emotion transfer system, called ET-GAN, which can synthesize emotional speech from any input speech for given emotion categories without fine-tuning TTS system. In order to synthesize purer emotion information by constraining the change of speaker characteristics, ET-GAN learns the mapping function from the input speech to the target speech without aligning the time sequences and speech samples. Meanwhile, due to emotional commonality in different languages, the emotion information learned from one language can be quickly applied to another language through transfer learning. The objective evaluation and the subjective evaluation shows that the emotional speech synthesized by ET-GAN has more superior performance on emotion correctness and speaker retention than the existing methods \cite{kameoka2018stargan,kaneko2018cyclegan,kaneko2019cyclegan}. Both subjective and objective evaluation proved that the use of domain adaptive technology can effectively achieve the migration of emotion between different languages for the task of language-independent emotion transfer.

However, compared with the original speech, the synthesized emotional speech still has low distortion; the synthesized emotional speech has a tiny gap with the real emotional speech in emotion correctness. In future work, we plan to improve the accuracy and authenticity of emotional speech synthesis by optimizing preprocessing methods, model structure, and training strategies. It is predictable that the proposed model can be applied not only to the task scenarios presented in this paper but also to other interest directions.

\ack
This work was partially supported by the National Key Research and Development Program of China (2016QY04W0903), Beijing Municipal Science and Technology Project (Z191100007119010) and National Natural Science Foundation of China (NO.61772078). 
Any opinions, findings, and conclusions or recommendations expressed in this material are those of the authors and do not necessarily reflect the views of any funding agencies.

\bibliography{ecai}
\end{document}